\documentclass[a4paper, reprint,twoside]{revtex4-2}
\usepackage[english]{babel}
\usepackage{xcolor}
\usepackage{ulem}
\usepackage{amsthm}
\usepackage{mathtools}
\usepackage{physics}
\usepackage{graphicx}
\usepackage[left=15mm,right=15mm,top=20mm,columnsep=15pt]{geometry} 
\usepackage{adjustbox}
\usepackage{placeins}
\usepackage[T1]{fontenc}
\usepackage{lipsum}
\usepackage{csquotes}
\begin{document}
\title{Active volatile drops on liquid baths}

\author{Benjamin Reichert$^{1}$, Jean-Benoît Le Cam$^{1}$, Arnaud Saint-Jalmes$^{1}$, Giuseppe Pucci$^{1}$ }
\email[Correspondence email address: ]{benjamin.reichert90@gmail.com;}
\affiliation{$^1$Univ Rennes 1, CNRS, IPR (Institut de Physique de Rennes)  UMR 6251, FR35000 Rennes, France}




\begin{abstract}
In most experimental studies, active drops propel in a liquid bulk due to self-generated interfacial stresses of solutal origin. Here, we demonstrate the self-propulsion of a volatile drop on the surface of a liquid bath due to stresses of thermal origin. Evaporative heat pumping is converted into directed motion driven by thermocapillary stresses, which emerge on the drop surface as a result of a symmetry breaking of the drop temperature field. The dependence of the drop speed on the activity source, i.e. the evaporation flux, is derived with scaling arguments and captures the experimental data. 
\end{abstract}


\maketitle

Active particles are self-propelled particles that convert ambient or internal energy into directed motion  \cite{ramaswamy2010mechanics,marchetti2013,bechinger2016,schweitzer2003}. In a fluid medium, squirmers are active particles that move in response to tangential stresses on their surface  \cite{lighthill1952,herminghaus2014,bechinger2016}. Squirmers can be natural, as some bacteria  \cite{blake1971,ehlers1996,drescher2009}, or artificial, as colloids \cite{bickel2013,popescu2018} or drops \cite{thutupalli2011,seemann2016}. Among artificial squirmers, particular attention has been devoted to active drops, which develop tangential stresses as a result of surface-tension gradients \cite{maass2016,ryazantsev2017,izri2014}. Since these drops are symmetric, spontaneous motion is triggered by a symmetry breaking of the particle interaction with its environment through the flow fields \cite{john2008,de2013,tjhung2012}. In most experimental studies, these drops are stabilized by a suitable surfactant that sustains solutal surface tension gradients through specific chemical reactions \cite{thutupalli2011,toyota2006,peddireddy2012}. 
 However, these gradients are difficult to probe and direct probing of interfacial stresses can be essential to unravel the fundamental mechanisms of self-propulsion and collective phenomena in active emulsions \cite{herminghaus2014}.  Liquid/air interfaces are suitable platforms for the self-propulsion of natural \cite{bush2006walking} and artificial bodies, including solid particles \cite{snezhko2009self,chung2009electrowetting,suematsu2010collective,karasawa2014simultaneous,grosjean2015remote}, drops \cite{couder2005walking,pucci2011mutual,pucci2015faraday,ebata2015swimming,bormashenko2015self} and robots \cite{yuan2012bio,chen2018controllable}, and enable direct probing of a number of physical quantities \cite{bush2006walking,snezhko2009self,karasawa2014simultaneous,ebata2015swimming}. 
 
 In this Letter, we experimentally demonstrate the self-propulsion of volatile drops floating at a liquid/air interface, characterize the temperature field on the drop surface, the evaporation and the hydrodynamic flows in the bulk, and rationalize with scaling arguments the dependence of the drop speed with the activity source, i.e. the evaporation flux. 

\begin{figure}[h!]
\centering
\includegraphics[width=0.999\linewidth]{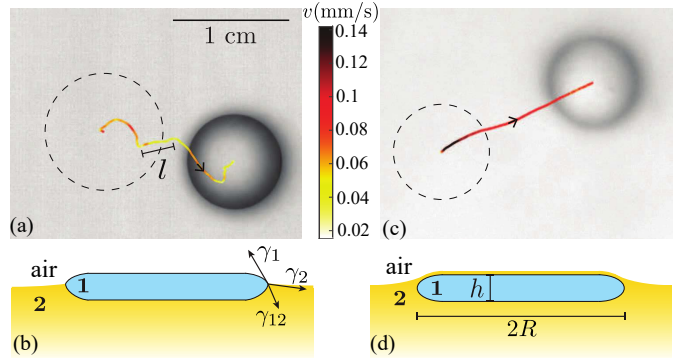}
\caption{The two archetypes of self-propulsion of a volatile drop on a liquid bath. (a, b) First archetype. (c, d) Second archetype. (a) The drop exhibits an erratic trajectory with typical persistence length $l\lesssim R$ (top view). The drop is in contact with air ($S_{21}<0$). (b) Schematic side view  (not to scale). (c) The drop exhibits a straight trajectory with $l\gg R$ (top view). A film of liquid 2 covers the drop ($S_{21}>0$). (d) Schematic side view (not to scale). (a) Drop of 90$\%$v/v ethanol in water. (c) Drop of ethanol. In (a,c), the drop initial volume is $V=0.1$ ml and liquid 2 is silicone oil with $\eta_2=0.097$ Pa.s.}
\label{fig1}
\end{figure}

\indent The experimental system consists of a drop of volatile liquid (liquid 1) floating on a bath of immiscible liquid (liquid 2). We tested a number of liquids for the drop, including ethanol, isopropanol, dichloromethane, and for the bath, including fluorinated and silicone oils. In the following we denote $\rho_1 (\rho_2)$, $\gamma_1 (\gamma_2)$ and $\eta_1(\eta_2)$ the density, surface tension with air and dynamic viscosity of liquid 1 (liquid 2), respectively. $\gamma_{12}$ is the interfacial tension between liquid 1 and liquid 2. We deposited drops with volume in the range $V=[0.1,0.4]$~ml, which adopted a pancake-like shape with radius $R\gg h$, where $h$ is the drop thickness (Fig. \ref{fig1}). The drop shape is the result of the dominant effect of gravity with respect to capillarity as  $R\gg l_{c\,1},l_{c\, 12} $ \cite{2013capillarity}, where $l_{c\,1}=\sqrt{\gamma_1/\rho_1\,g}$ and $l_{c\,12}=\sqrt{\gamma_{12}/(\rho_2-\rho_1)\,g}$  are the capillary lengths associated to liquid 1 and to the interface between liquid 1 and liquid 2, respectively, and $g$ is the gravitational acceleration. The drop equilibrium thickness is given by the Langmuir prediction \cite{langmuir1933} $h=\sqrt{-2\,S_{12}\,\rho_{2}/\rho_{1}(\rho_{2}-\rho_{1})\,g  } $, where $S_{12}=\gamma_{2}-(\gamma_{12}+\gamma_{1} )<0$ is the spreading parameter of liquid 1 on liquid 2. For the pairs of liquids considered in this study, $h\sim 1$ mm and $R\sim 1$ cm. 
The depth of the liquid bath was fixed to $H=10$ cm in all experiments. For all the liquid pairs we tested, we observed that the drop initially stays still for a time $\sim 1$ minute, then spontaneously sets into motion with speed $v\sim 0.01 - 0.1$~mm.s$^{-1}$. 

We identified two archetypes of behavior differing in the features of the drop trajectory. In the first archetype the drop undergoes an erratic motion characterized by a persistence length $l \lesssim R$ (Fig. \ref{fig1}(a)). In the second archetype, the drop trajectory is significantly straighter with $l\gg R$, and the drop achieves a stationary speed after a transient acceleration phase  (Fig. \ref{fig1}(c)). We found that straight trajectories with stationary speed occur when a thin film of liquid 2 covers the drop. This film is absent for liquid pairs exhibiting the first archetype of behavior.
The presence of the film is due to a peculiar wetting configuration called pseudo-total wetting \cite{sebilleau2013}, for which the spreading parameter of liquid 2 on liquid 1 is positive, $S_{21}=\gamma_{1}-(\gamma_{12}+\gamma_{2} )>0 $, while $S_{12}=\gamma_{2}-(\gamma_{12}+\gamma_{1})<0$. We investigate the mechanism underlying self-propulsion by focusing on the second archetype of behavior and we choose the pair ethanol/silicone oil as the representative pair of this archetype (fluid properties in Sec. 1 in Supplemental Material). 

\begin{figure}[h]
\centering
\includegraphics[width=0.999\linewidth]{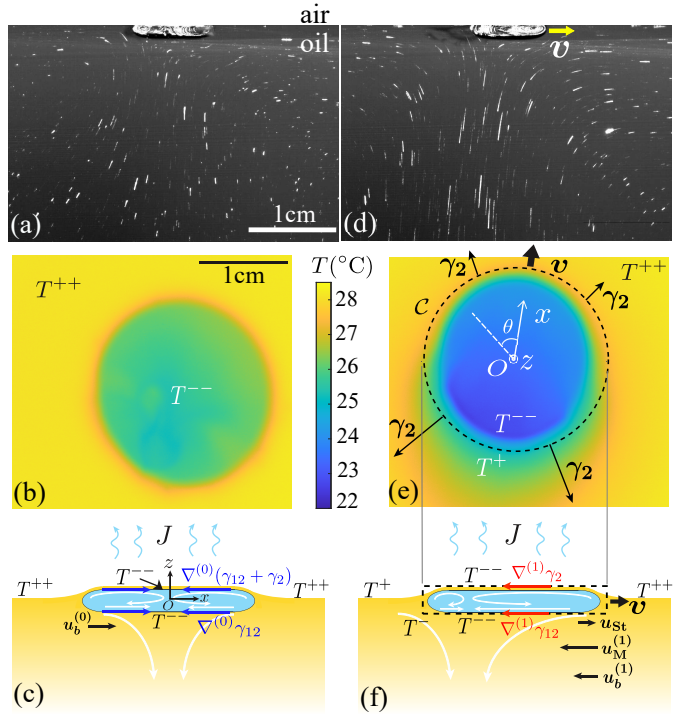}
\caption{Static symmetric and propulsive asymmetric state for a drop of ethanol on a bath of silicone oil. (a-c) The drop is at rest. (d-f) The drop moves with speed $v$.  (a,d) Experimental flows captured by particle tracking velocimetry by tracking each particle for 15 seconds (side view).  (b,e) Thermal images of the surface temperature field (top view), with $T^{++}>T^{+}>T^{-}>T^{--}$. (c,f) Side view schematics (not to scale). In (e,f) the dashed line delimits the imaginary box on which the force balance is performed. Bath viscosity is $\eta_2=0.097$ Pa.s. Drop volume is  $V=0.1$ ml in (a,d) and $V=0.4$ ml in (b,e).}\label{fig2}
\end{figure}

 Drop destabilisation and self-propulsion result from continuous conversion of evaporation heat into liquid motion. In order to understand the mechanism leading to propulsion, thermal imaging and particle tracking velocimetry (PTV) (see Sec. 3 and Sec. 4 in Supplemental Material) were used to characterize the system temperature and velocity fields during the static symmetric (Fig. \ref{fig2}(a-c)) and steady propulsion (Fig. \ref{fig2}(d-f) phases for $\eta_2=0.097$ Pa.s. Once the drop is deposited on the bath, the film develops on its surface within $\sim 100$ ms and ethanol pervaporates by diffusing through the film (Fig. \ref{fig2}(c)). The drop cools down and displays axisymmetric flow and temperature fields (Fig. \ref{fig2}(a,b)), with average temperature gradient $(T^{++}-T^{--})/R>0$ in the radial direction (Fig. \ref{fig2}(b,c)).  Two centripetal flows develop in the vicinity of the upper and lower interfaces and a centrifugal flow develops in the bulk of the drop (Fig. \ref{fig2}(a,c)). Flow reversal occurs in the vicinity of the drop's axis of symmetry ($Oz$), which is characterized by a vanishing radial component of the velocity. In the oil, the flow is centripetal in the vicinity of the drop's lower interface and recirculates outwardly towards the bottom of the bath over typical distance $\sim R$ (Fig. \ref{fig2}(a,c)). These flows are driven by interfacial stresses arising from the variation of interfacial tension with temperature along the radial direction. These thermal Marangoni stresses are denoted $\vb*{\nabla}^{(0)}( \gamma_{12} + \gamma_{2})$ and  $\vb*{\nabla}^{(0)} \gamma_{12}$ for the upper and lower interfaces, respectively, and are directed toward the axis of symmetry of the drop (Fig. \ref{fig2}(c)). Here, the index $(0)$ refers to the flow generated by these stresses, which remains the base flow throughout the whole dynamics of the drop. 

\begin{figure}[h!]
\centering
\includegraphics[width=0.999\linewidth]{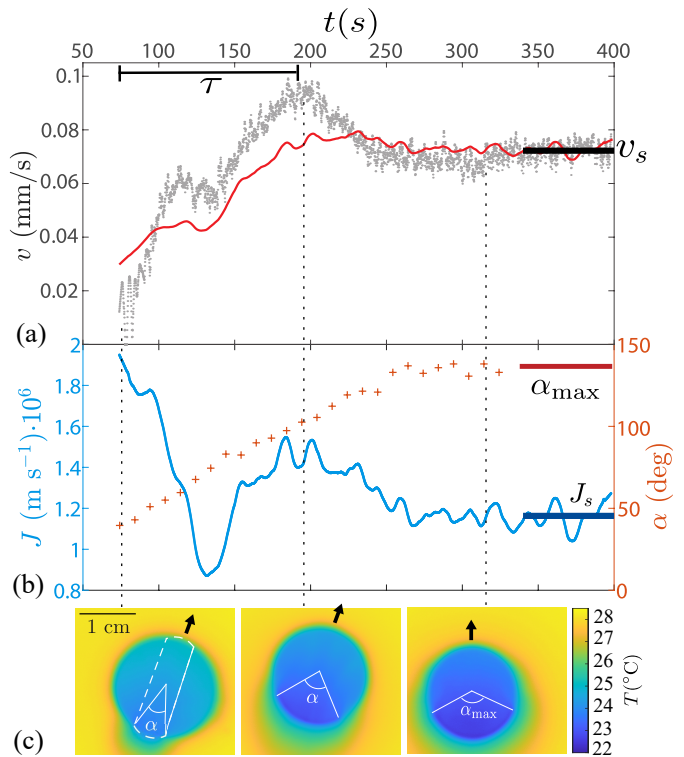}
\caption{Experimental time evolution of drop speed, evaporation flux and temperature field. (a) Drop speed $v(t)$ ($\textcolor{gray}{*}$).  $v_s$ is the stationary speed. The solid line (\textcolor{red}{-}) results from \eqref{eq_gene_v_s_temp} combined with the evaporative flux $J(t)$ plotted in (b). (b) Evaporation flux $J(t)$ ($\textcolor{blue}{-}$) and aperture angle $\alpha(t)$ ($\textcolor{red}{+}$)  of  the crescent-like cold patch. $J_s$ is the stationary evaporation flux and $\alpha_{\rm max}$ is the maximum angle. (c) Thermal images showing the temporal evolution of the surface temperature field. The angular extension $\alpha$ of the crescent-like patch increases with time. In (c) the dashed white line delimits the upper side of the imaginary box on which the force balance is performed in the transient regime. The drop volume is $V=0.4$ ml and $\eta_2=0.097$ Pa.s.}
\label{fig3}
\end{figure}

\noindent About one minute after the drop deposition, thermal images reveal a symmetry breaking of the temperature field. A crescent-like cold patch appears in the vicinity of the drop horizontal contour (Fig. \ref{fig3}(c)), the angular extension of which, $\alpha$, increases with time until it achieves a stationary value $\alpha_{\rm max}$. We denote with $T^{--}$ the average temperature of the patch in this asymmetric state. PTV images show that the zone of flow reversal shifts accordingly (Fig. \ref{fig2}(d,f)). Correspondingly, the drop starts moving in the direction opposite to the displacement of the patch, along the direction of the average temperature gradient ($\mathbf{\hat{x}}$ direction in Fig. \ref{fig2}(e,f)). 

In order to rationalize the mechanics of self-propulsion, we note that the Reynolds number comparing advection to viscous transport of momentum in the bath is $Re=\rho_2 v \,R/\eta_2\sim 10^{-2}$ with $\eta_{2}=$0.1 Pa.s. We thus use the Stokes equation and describe the global flow during the stationary propulsion (Fig. \ref{fig2}(d,f)) as the superposition of an order $(0)$ symmetric flow and an order (1) perturbation flow. The order $(1)$ flow is associated to the average temperature difference $\Delta T=T^{++}-T^{+}>0$ between the fore and the aft of the drop. $\Delta T$ induces Marangoni stresses on the upper and lower interfaces that drag liquid from the fore to the aft, and are responsible for the displacement of the region of flow reversal (Fig.~\ref{fig2}(f)). 

\noindent In order to derive a scaling law for the drop stationary speed, we consider the $x$-component of the forces experienced by an imaginary, cylindrical control volume of radius $R$ and thickness $h$ containing the drop (dashed box in Fig.~\ref{fig2}(e,f)). These forces arise as a result of the perturbation flow outside the drop, which we consider as the sum of two flows (Fig.~\ref{fig2}(f)). We develop our model in the laboratory frame of reference. \\
The first flow is the Stokes flow associated to the  motion of a viscous disc in a quiescent fluid without Marangoni effect \cite{hadamard1911,rybczynski1911}. 
This results in the Stokes viscous drag $F_{\rm St} \propto -\eta_2\, R \,v_s < 0$ \cite{guyon2001}, with a prefactor assumed as constant as it weakly depends on $h/R$ and $\eta_1/\eta_2$ for $(h/R),(\eta_1/\eta_2) \ll 1$ \cite{kimkarrila2013, hadamard1911, rybczynski1911}. We denote $u_{\rm St}>0$ the characteristic velocity of the Stokes flow in the $\mathbf{\hat{x}}$ direction close to the lower interface in liquid 2 (Fig.~\ref{fig2}(f)). \\ 
The second flow is due to the Marangoni stresses at the interface of a drop that is at rest with respect to the bath. We denote $\vb*{\nabla}^{(1)}\gamma_{2}$ and $\vb*{\nabla}^{(1)} \gamma_{12}$ the gradients at the upper and lower interfaces, respectively, which generate two forces. 
A force is the result of the inhomogeneous, radial and outwardly pulling tension $\vb*{\gamma_2}[T(\theta)]$, 
which is due to azimuthal temperature variations. It applies to the closed contour $\mathcal{C}$ (Fig. \ref{fig2}(e)) and writes 

\begin{equation}
F_{\rm cam}=\oint_{\mathcal{C}} \vb*{\gamma_2}\cdot\vb*{\mathrm{\hat{x}}} \,dl=\int_{\mathcal{S}}\vb*{\nabla}^{(1)}\gamma_2\cdot\vb*{\mathrm{\hat{x}}} \, dS \sim - \frac{|d\gamma/d T|\Delta T}{R}\cdot \pi R^2, 
\end{equation}
where $\mathcal{S}$ is the area of the upper side of the control volume (Fig. \ref{fig2}(e)). Here and in the following we assume $|d\gamma/d T|\sim |d\gamma_2/d T|\sim|d\gamma_{12}/d T|$ (\cite{girifalco1957} and Sec.1 in Supplemental Material) . 
This force is the thermal analog of the force that drags camphor boats towards zones with higher interfacial tension \cite{nakata2005,suematsu2010collective}, but here this force opposes the drop motion, $F_{\rm cam}<0$.\\
A second force arises as a result of the Marangoni stress $\vb*{\nabla}^{(1)}\gamma_{12}$ on the drop's lower interface, which induces a viscous stress response in the underlying bath that is oriented in the direction of motion of the drop. In other words, the drop gains traction from the bath in order to propel. We denote $u_{\rm M}^{(1)}<0$ the characteristic velocity of the perturbed flow in the $\mathbf{\hat{x}}$ direction close to the lower interface in liquid 2 (Fig. \ref{fig2}(f)). This flow produces a strain rate $\sim |u^{(1)}_{\rm M}|/R$ \cite{schmitt2016}  and the resulting force on the lower side of the control volume writes
\begin{equation}
F_{\rm prop}\sim\,\eta_2 \frac{\vert{u_{\rm M}^{(1)}\vert}}{R} \cdot \pi R^{2}. 
\label{eqProp}
\end{equation}
\noindent The problem is thus reduced to computing the scaling of $u_{\rm M}^{(1)}$. In our experiments, the viscous stress contribution inside the drop is negligible with respect to the contribution from the outer liquid, $\eta_1/h\ll\eta_2/R$. As a result, the continuity of stresses on the lower interface yields $u_{\rm M}^{(1)} \propto - |d\gamma/dx| R/\eta_2 \sim - |d\gamma/d T| \Delta T/\eta_2$ (details in Sec. 5 in Supplemental Material). Substituting this expression in \eqref{eqProp} we obtain  $F_{\rm prop} \propto |d\gamma/d T| \Delta T R > 0$. We note that $F_{\rm prop}$ has the same scaling as $F_{\rm cam}$, but opposite direction. As the drop is moving in the direction of $F_{\rm prop}$, we may write $F_{\rm prop} + F_{\rm cam} \propto |d\gamma/d T|\Delta T R > 0$. The stationary drop speed scaling can thus be obtained by balancing this result with $F_{\rm St}$, which yields
\begin{equation}
\centering 
v_s\propto \left|\frac{d\gamma}{dT} \right| \frac{\Delta T}{\eta_2}.  
\label{eq1}
\end{equation}

\indent In order to derive a scaling for $\Delta T$, we analyze how the evaporation energy is converted into liquid motion. We simultaneously measured the drop speed $v(t)$, the angular extension of the cold patch $\alpha(t)$ and the evaporation flux $J(t)$, which drives the drop dynamics (Fig. \ref{fig3}). In this experiment, the drop is deposited on the bath at $t=0$~s  and starts moving  at $t=74$~s. After a transient regime of acceleration with characteristic time $\tau\sim 100$ s, the drop achieves a stationary speed $v_s$. We define $J(t)=-dV(t)/dt / A(t)$, where $A(t)=\pi\, R^{2}(t)$ is the area and  $V(t)=h\,A(t)$ is the volume of the drop. We measured $A(t)$ from top-view video-recording with an in-house algorithm (Sec. 6 in Supplemental Material). $v(t)$, $\alpha(t)$ and $J(t)$ exhibit the same time scale of evolution toward a stationary value, which suggests that $v(t)$ is strongly correlated to both $\alpha(t)$ and $J(t)$.\\ 
We focus on the stationary regime and calculate how $v_s$ scales with the thermal evaporation flux, $\rho_1\, \mathcal{L}_v \,J_s$, where $\mathcal{L}_{\rm v}$ 
is the latent heat of evaporation of ethanol. A thermal flux balance is performed to relate $\Delta T$ to the evaporative heat flux. $u_{b}^{(1)}=u_{\rm M}^{(1)} + u_{\rm St} < 0$ is the characteristic velocity of the total perturbed flow in the $\mathbf{\hat{x}}$ direction. Heat transport in the bath is governed by two processes, convection and diffusion, the relative importance of which is quantified by a thermal Péclet number for each component of the flow, $Pe^{(0)}= R\,u^{(0)}_{b}/D_{\rm th}$ and $Pe^{(1)}= R\,u^{(1)}_{b}/D_{\rm th}$. $D_{\rm th}=\lambda_2/\rho_2\,C_{\rm p}$ is the thermal diffusion coefficient, with $C_{\rm p}$ and $\lambda_2$ the thermal capacity and the thermal conductivity of silicone oil, respectively. For $R\approx 8$ mm and measured velocities  $u^{(0)}_{b}\approx0.13$ mm.s$^{-1}$ and $u^{(1)}_{b}\approx0.05$ mm.s$^{-1}$ for $\eta_2=0.097$ Pa.s, we obtain $Pe^{(0)}\approx 9$ and $Pe^{(1)}\approx 4$, which suggest that convective transport is dominant. The temperature difference $\Delta T$ is sustained by the cooling of the fluid particles that move backward in the vicinity of the drop's upper interface, driven by the thermal evaporative flux $\rho_1 \mathcal{L}_{\rm v}\, J_s$. This temperature difference induces thermal convection in the underlying bath, $\rho_2\, C_{\rm p} \,\Delta T \,\vert u_{b}^{(1)} \vert$, from the fore towards the aft of the drop. The heat flux balance thus writes
\begin{equation} \rho_2\, C_{\rm p} \,\Delta T \,\vert u_{b}^{(1)} \vert=c\, \rho_1 \mathcal{L}_{\rm v}\, J_s \label{ordre_1},\end{equation}
where $c$ represents the fraction of thermal evaporation flux that is converted into the convective heat flux of the perturbed flow. By substituting $u^{(1)}_{b}\approx0.05$ mm.s$^{-1}$, $J_s=1.2 \times 10^{-6}$\, m.s$^{-1}$ and $\Delta T\simeq 1$~$^{\circ}$C from experiments in \eqref{ordre_1}, we found $c=0.09$, which suggests that the evaporative flux is not modified by the drop propulsion at leading order.

\noindent In order to determine a scaling for $\Delta T$, we note that $u_{\rm St}\sim v_s$ for the Stokes flow past a viscous body in the limit $\eta_1/\eta_2\ll1$ \cite{guyon2001}. Since $u_{\rm M}^{(1)}$ and $v_s$ have the same scaling (cf. \eqref{eq1}) then $u^{(1)}_{b}\propto - |d\gamma/d T|\,\Delta T/\eta_2 $. We thus obtain $\Delta T\propto \left[\eta_2\,\rho_1\,\mathcal{L}_v\,J_s / \rho_2\,C_{\rm p}\,|d\gamma/dT|\right]^{1/2}$, that combined with \eqref{eq1} yields the scaling of the drop stationary speed with the evaporation flux $v_s\propto \left[\rho_1 \,|d\gamma/dT|\,\mathcal{L}_v\, J_s / \rho_2\,\eta_2\, C_{\rm p} \right]^{1/2}.$

We now focus on the transient regime, which is characterized by an increasing angular extension $\alpha(t)$ of the cold crescent over the typical time $\tau\approx 100$~s (Fig.~\ref{fig3}(b,c)). Since $\tau$ is much larger than the typical time scale of viscous diffusion of momentum in the bath, $\tau_v = \rho_2\,R^{2}/\eta_2\sim~1$~s, we consider the velocity field evolution in the bath as quasi-stationary. Consequently, the scaling of $v_s$ can be extended to the transient regime by considering an appropriate geometrical factor $g[\alpha(t)]$ that takes into account the time evolution of the cold crescent-like patch (Fig. \ref{fig3}(c)). During the transient regime, $\nabla^{(1)}\gamma_{2}$ and $\nabla^{(1)} \gamma_{12}$ apply to the area $S_{\rm tr}(\alpha)=2\,g(\alpha)\,R^2$ at the upper and lower side of a time-varying control volume, respectively, where $g(\alpha)=\alpha/2+ \cos(\alpha/2)\sin(\alpha/2)$. $S_{\rm tr}$ is a fraction of the total drop surface $\pi R^2$ and its contour is represented by a white dashed line in Fig. \ref{fig3}(c). We perform the force balance on the time-varying control volume and we obtain
\begin{equation}
    v(t) = \beta g(\alpha) \left[\frac{\rho_1 \,|d\gamma/dT|\,\mathcal{L}_v\, J(t)}{\rho_2\,\eta_2\, C_{\rm p}} \right]^{1/2}.
\label{eq_gene_v_s_temp} 
\end{equation}
Using the experimental flux $J(t)$ in Fig.~\ref{fig3}(b) this expression yields a reasonable fit to the experimental speed with $\beta=0.09$ (solid red line in Fig.~\ref{fig3}(a)). For $\alpha = \alpha_{\rm max}$ we recover the stationary regime.

\begin{figure}[t!]
\centering
\includegraphics[width=0.999\linewidth]{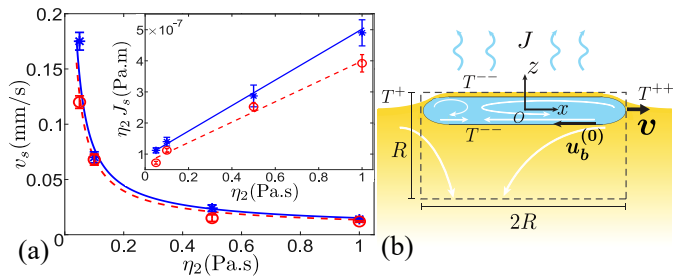}
\caption{Relation between the drop speed and the evaporation flux in the stationary regime of self-propulsion. (a) Stationary speed of the drop $v_s$ as a function of the bath viscosity for $V=0.1$ ml (\textcolor{blue}{$\star$}) and $V=0.4$ ml (\textcolor{red}{o}). Lines represent the scaling for $v_s$ from \eqref{eq_final} combined with the fits in the inset. $V=0.1$ ml (continuous line), $V=0.4$ ml (dashed line). Inset: Dependence of $\eta_2\,J_s$ on the bath viscosity $\eta_2$ for $V=0.1$ ml (\textcolor{blue}{$\star$}) and $V=0.4$ ml (\textcolor{red}{o}) . Lines represent fits of $a\,\eta_2 +b$ to each set of data with $(a,b)=(4.1 \times 10^{-7} \,\textrm{m.s}^{-1}, 9.2\times 10^{-8}\, \textrm{m.Pa} )$ for $V=0.1$ ml (continuous line) and $(a,b)=(3.3\times 10^{-7}\,\textrm{m.s}^{-1} ,7.1\times10^{-8}\, \textrm{m.Pa})$ for $V=0.4$ ml (dashed line). (c) Side view schematic of the drop in motion (not to scale). The dashed rectangle represents the box used for the thermal balance. }
\label{fig4}
\end{figure}

In order to determine the physical mechanism that underlies $J(t)$, we compare the typical time scales of the transport phenomena that may limit its magnitude. The time for diffusion of ethanol through the film is $t^{\rm film}_{\rm diff} = h_f^2/D_{\rm 12}\sim0.001$~s, where $D_{12}\sim10^{-9}$ is the mass diffusion coefficient of ethanol in silicon oil \cite{chuan2011measurement} and $h_f\sim300$~nm is the film thickness (Sec. 2 in Supplemental Material). The time for diffusion of ethanol in air is $t^{\rm air}_{\rm diff}\sim R^2/D_{\rm 1} \sim 10$~s, where $D_{\rm 1}\sim10^{-5}$ m$^2$.s$^{-1}$ is the mass diffusion coefficient of ethanol in air \cite{fukatani2016effect}. These are both smaller than the time for thermal convection and diffusion in the bath, $t^{ \rm bath}_{\rm conv}\sim R/u^{(0)}_{b}=100$ s and $t^{\rm bath}_{\rm diff}\sim R^2/D_{\rm th}=1000$~s, respectively. As a result $J(t)$ is limited by thermal convection in the bath. We thus proceed by varying the bath viscosity $\eta_2$, which affects the advective thermal transport in the bath and thus the activity of the drop. We measured the variation of stationary evaporation flux and drop speed with $\eta_2$ for drops with two different volumes $V=0.1$~ml ($R=4.0$~mm) and $V=0.4$~ml ($R=8.0$~mm) (Fig. \ref{fig4}).  $\eta_2\,J_s$ is an affine function of $\eta_2$ (Fig. \ref{fig4}(a)), thus $J_s=a+b/\eta_2$. We identify $a$ as the diffusive term and $b/\eta_2$ as the convective term, the decoupling of which is typical for the heat transfer from a body immersed in a flow with low $Re$ and $Pe$ up to ${\cal{O}}(1)$ \cite{acrivos1962}. In our system, this decoupling appears to be still valid for measured $Pe^{(0)}=[2,11]$. The diffusive contribution $a$ is estimated by the formula for the diffusion flux from a disk \cite{crank1979}, $\rho_1 \mathcal{L}_{\rm v}\,a=4 \,\lambda_2 (T^{++}-T^{--})/\pi R$ at temperature $T^{--}$ in a medium at temperature $T^{++}$. The convective term $b/\eta_2$ can be estimated from the convective thermal flux associated to the order $(0)$ flow in the bath. A scaling for $b/\eta_2$ is obtained with a thermal balance in a two-dimensional box containing the drop and the underlying bath (Fig. \ref{fig4}(b)). Hot liquid in the bath enters the box through its sides with temperature $T^{++}$ and velocity $u^{(0)}_{b}$, while cold liquid with temperature $T^{--}$ exits the box through its lower side with the same speed.  Heat is pumped through the upper side of the box with flux $\rho_1\,\mathcal{L}_{\rm v} \,b/\eta_2$. As a result, $\rho_1\,\mathcal{L}_{\rm v} \,b/\eta_2=\rho_2\,C_{\rm p} (T^{++}-T^{--})\,u^{(0)}_b$. For $(T^{++}-T^{--})=5$ K, $R=8.0$ mm and $u^{(0)}_{b}=0.13$ mm.s$^{-1}$ for $\eta_2=0.097$ Pa.s this approach yields $a=1.9\times 10^{-7}$m.s$^{-1}$ and $b=1.4\times 10^{-7}$ m.Pa, which are in the same order of the values from the experimental fit (inset in Fig. \ref{fig4}(a)). The combination of the fit $J=a+b/\eta_2$ with the scaling for the drop stationary speed in \eqref{eq_gene_v_s_temp} with fixed $\beta=0.09$ and $g(\alpha_{\rm max})$ yields
\begin{equation}v_s=\beta\, g(\alpha_{\rm max}) \left[\frac{\rho_1 \,|d\gamma/dT|\,\mathcal{L}_v\, (a+b/\eta_2)}{\rho_2\,\eta_2\, C_p} \right]^{1/2}, \label{eq_final}
\end{equation} 
which captures well the dependency of the stationary speed on the bath viscosity (Fig. \ref{fig4}(a)). 

The symmetry breaking of the temperature field results from an instability \cite{ryazantsev2017}. We now discuss the conditions for the instability to occur in our system. The temperature difference $\Delta T$ is amplified by the cooling of ethanol fluid particles at a rate $\mathcal{L}_v \,J$ as they are convected backwards in the vicinity of the drop upper interface. The characteristic velocity $V_{\rm M}$ of this Marangoni convection is given by the continuity of stress on the same interface $\eta_1\, V_{\rm M}/h\sim |d\gamma/d T|\,\Delta T $. On the other hand, in our system $\lambda_2 \sim \lambda_1$, therefore we denote with $\lambda$ the thermal conductivity of both liquids. Thermal diffusion in the negative $x$-direction mitigates the temperature discrepancy as the drop starts moving. The thermal Péclet number that drives the instability can be derived by considering $V_M$ and the balance between the stabilizing, diffusive thermal flux and the thermal evaporation flux, $ \lambda\, \Delta T/R\propto \rho_1\mathcal{L}_v \,J$. We thus obtain  
\begin{equation}Pe=R\, V_M /D_{\rm th}\propto \rho_2 C_p R
^2\,h\left|\frac{d\gamma}{d T}\right| \frac{\rho_1 \mathcal{L}_v\, J}{\eta_1\, \lambda^2}.\label{pe}\end{equation}

\noindent This expression is similar to the Péclet number governing the drift instability of a hot drop cooling down in a liquid bulk with uniform temperature \cite{rednikov1994}. Stability analysis indicates that the instability occurs above a threshold value of this non-dimensional parameter, for which heat convection dominates diffusion and the drop starts moving \cite{rednikov1994}. While \eqref{pe} includes the relevant parameters required for self-propulsion to occur in our system, the analog threshold value of the Péclet number remains unknown.


\indent We have reported the first experimental evidence of self-propulsion of a volatile drop on a liquid bath due to thermocapillary stresses. Two archetypes of propulsion are identified, differing in both the presence of a film coating the drop and the persistence length of the drop trajectories. The drop motion occurring at a liquid/air interface, we could probe the surface temperature and relate thermal stresses to hydrodynamic flows. While the presence of the film does not limit the evaporation flux, thermal imaging of both archetypes reveales that the film substantially attenuates surface temperature fluctuations (Sec. 7 in Supplemental Material), which are a known feature of evaporating sessile drops \cite{sefiane2008} and are responsible for the erratic nature of the drop motion. We characterized the archetype yielding straighter trajectories in order to elucidate the physical mechanism underlying self-propulsion. 

Self-propulsion is triggered by a thermo-capillary convective instability \cite{ryazantsev2017} and associated with the emergence of a propulsive force exerted on the drop by the Marangoni stresses on its lower interface. The force results from convection-sustained temperature gradients along the drop interface, resulting in a warmer pool of liquid being advected by the hydrodynamic flow in the underlying bath toward the back of the drop.\\
Although the drop is located at an interface and thereon experiences Marangoni stresses, its dynamics differ from a solid Marangoni surfer \cite{nakata2005,suematsu2010collective,lauga2012,wurger2014} because the drop shares a fluid interface with the bath. The Marangoni stresses occurring at this interface yield a propulsion scheme that is rather similar to a classical squirmer \cite{lighthill1952,herminghaus2014} that gains traction from the external medium in order to move. This results into a direction of motion opposite to the interfacial tension gradient and thus to the direction of motion a classical Marangoni surfer. 

\section*{Acknowledgements} \label{sec:acknowledgements}
   B.R. and G.P. thank the program CNRS Momentum 2017 for its support. The authors thank Isabelle Cantat and Adrien Bussonnière for useful discussions.



\bibliography{active_droplet}

\appendix*
\section*{Supplemental Material}
\subsection{Liquids properties} Ethanol has density $\rho_1=790$ Kg.m$^{-3}$, dynamic viscosity $\eta_1=1.2\times 10^{-3}$ Pa.s and latent heat of evaporation $\mathcal{L}_{\rm v}=8.55\times 10^{5} $ J.Kg$^{-1}$  (VWR AnalaR NORMAPUR $\geq 99.8$ $\%$). The silicone oils (RHODORSIL) we used have density $\rho_2=970$ Kg.m$^{-3}$, dynamic viscosities $\eta_2=\left\{0.0485,0.097,0.485, 0.97\right\}$ Pa.s, thermal conductivity $\lambda_2=0.16$~J.s$^{-1}$.m$^{-1}$.K$^{-1}$ and thermal capacity $C_{\rm p}=1460$~ J.Kg$^{-1}$.K$^{-1}$. Interfacial tensions $\gamma_1=22\pm 0.1$ mN.m$^{-1}$, $\gamma_2=20.6\pm 0.1$ mN.m$^{-1}$ and $\gamma_{12}=0.8\pm 0.1$ mN.m$^{-1}$ were measured using a pendant drop apparatus. All reported values are at $T=20$ $^{\circ}$C. \\
In this article, we assume $|d\gamma/d T|\sim |d\gamma_2/d T|\sim|d\gamma_{12}/d T|$ since $|d\gamma_{12}/d T|$ is in the same order than $|d\gamma_2/d T|=5\times 10^{-5}$ N.m$^{-1}$.K$^{-1}$ for silicone oil and  $|d\gamma_1/d T|=8\times 10^{-5}$ N.m$^{-1}$.K$^{-1}$ for ethanol \cite{girifalco1957}.

\subsection{Measurement of the film thickness $h_f$} The film thickness at the center of the drop in the stationary regime was measured by interferometry using a spectrometer (Spectra) \cite{champougny2016}. The measurement yielded $h_f\sim300$~nm.

\subsection{Thermal imaging} A thermal camera (FLIR X6540sc InSb, 640 x 512 pixels, wavelength ranges between 1.5 and 5.1 $\mu$m, detector pitch of 15 $\mu$m) was used to image the system surface temperature. The thermal resolution, namely the noise-equivalent temperature difference (NETD), is equal to 20 mK at 25  $^{\circ}$C. The acquisition frequency was set to 5 Hz. 

\subsection{Particle tracking velocimetry} The flow velocity field was measured with a standard particle tracking velocimetry setup consisting of a vertical laser (wavelength $488$ nm, Oxxius) sheet ($Oxz$ plane), seeding particles (Hollow glass Microspheres, Cospheric) and a videocamera placed orthogonaly to the laser sheet.

\subsection{Justification of the scaling for $u^{(1)}_{\rm M}$}
We derive the order (1) Marangoni flow associated to the temperature difference $\Delta T$  assuming a static drop and a two dimensional flow in the plane $Oxz$ (Fig.~2(e,f)). Since $R\gg h$ the stationary flow inside the drop is quasi-parallel along the $x$ direction. We can thus use the lubrication approximation and neglect the convection terms in the Navier-Stokes equation governing the flow \cite{guyon2001}. Furthermore, the pressure and the velocity fields do not depend on $z$ and $x$, respectively. The flow obeys the Stokes equation

\[\frac{\partial^2 v_{1}^{(1)}}{\partial z^2}=\frac{1}{\eta_{1}}\frac{\partial P^{(1)} }{\partial x}=A, \]

\noindent from which 

\[v_1^{(1)}=\frac{A}{2} z^2 + B z + C, \] with coefficients $A,B,C$ independent of the $z$ coordinate.

\noindent The dynamic boundary conditions consist in the  continuity of tangential stresses at the upper and lower interfaces, located at $z=h/2$ and $z=-h/2$, respectively,

\[ \frac{\partial (\gamma_2+ \gamma_{12})^{(1)}}{\partial x}|_{z=h/2} = \eta_1\frac{\partial v_1^{(1)}}{\partial z}|_{z=h/2,}\]

\[ \eta_1 \frac{\partial v_1^{(1)}}{\partial z}|_{z=-h/2} + \frac{\partial \gamma_{12}^{(1)}}{\partial x}|_{z=-h/2}=\eta_2 \frac{\partial v_2^{(1)}}{\partial z}|_{z=-h/2}. \]

\noindent The kinematic boundary condition at the lower interface gives

\[v_1^{(1)}(z=-h/2)=\frac{A}{8} h^2 - B \frac{h}{2} + C=u^{(1)}_{\rm M}.\] 

\noindent These three boundary conditions allow to express the coefficients as a function of the Marangoni stresses, the viscosities and $u^{(1)}_{\rm M}$

\[\begin{aligned}A=&\ \frac{1}{\eta_1\,h}\left[\frac{\partial (\gamma_2 + \gamma_{12})^{(1)} }{\partial x}|_{z=h/2} + \frac{\partial \gamma_{12}^{(1)}}{\partial x}|_{z=-h/2}\right] \\
&\ - \frac{\eta_2}{\eta_1\,h} \, \frac{\partial v_2^{(1)}}{\partial z}|_{z=-h/2},\end{aligned}\]

\[\begin{aligned}B= &\ \frac{1}{2\,\eta_1} \left[ \frac{\partial (\gamma_2 + \gamma_{12})^{(1)} }{\partial x}|_{z=h/2} - \frac{\partial \gamma_{12}^{(1)}}{\partial x}|_{z=-h/2}\right]\\ 
&\ +  \frac{\eta_2}{2\,\eta_1} \, \frac{\partial v_2^{(1)}}{\partial z}|_{z=-h/2}  ,\end{aligned} \]

\[\begin{aligned}C= &\ u^{(1)}_{\rm M}+ \frac{h}{8\,\eta_1}\left[\frac{\partial (\gamma_2 + \gamma_{12} )^{(1)} }{\partial x}|_{z=h/2}-3 \frac{\partial \gamma_{12}^{(1)}}{\partial x}|_{z=-h/2}\right] \\
 &\ + \frac{3\,\eta_2\,h}{8\,\eta_1}\, \frac{\partial v_2^{(1)}}{\partial z}|_{z=-h/2}
 \end{aligned}\]

\noindent Mass conservation in the static drop writes

\[\int^{h/2}_{-h/2}v_1^{(1)} \,dz=\frac{A}{24}\,h^3+C\,h  =0\] and yields

\begin{equation}\begin{aligned}0=&\frac{\partial (\gamma_2 + \gamma_{12} )^{(1)} }{\partial x}|_{z=h/2}-2\,\frac{\partial \gamma_{12}^{(1)}}{\partial x}|_{z=-h/2} \\
&+ 2\,\eta_2\,\frac{\partial v_2^{(1)}}{\partial z}|_{z=-h/2}  + 6\,\eta_1\frac{u^{(1)}_{\rm M}}{h}   .\label{total}\end{aligned}\end{equation}

\noindent The viscous strain rate in the bath is $\frac{\partial v_2^{(1)}}{\partial z}|_{z=-h/2}\sim u^{(1)}_{\rm M}/R$ \cite{schmitt2016}. This scaling shows that the last term in \eqref{total} can be neglected for $(\eta_1/\eta_2)(R/H)\ll 1$, a condition that is verified in our experiments.

\noindent As $\frac{\partial (\gamma_2 + \gamma_{12} )^{(1)} }{\partial x}|_{z=h/2},\frac{\partial \gamma_{12}^{(1)}}{\partial x}|_{z=-h/2} \propto - |d\gamma/dT|\Delta T/R $, the scaling for $u^{(1)}_{\rm M}$ is

\[ u^{(1)}_{\rm M}\propto -|d\gamma/dT|\Delta T/\eta_2.\]

\subsection{Image Analysis: velocity and evaporation flux} The drop motion was recorded using a CMOS camera (Allied Vision Mako u-130B) with acquisition frequency set to 10 fps. Image processing to determine the experimental drop velocity and evaporation flux was performed with an in-house Matlab algorithm that detects the drop contour by performing image thresholding, and fitting the contour with an ellipse.

\subsection{Influence of the wetting film on the surface temperature}
\begin{figure}[h!]
\centering
\includegraphics[width=0.999\linewidth]{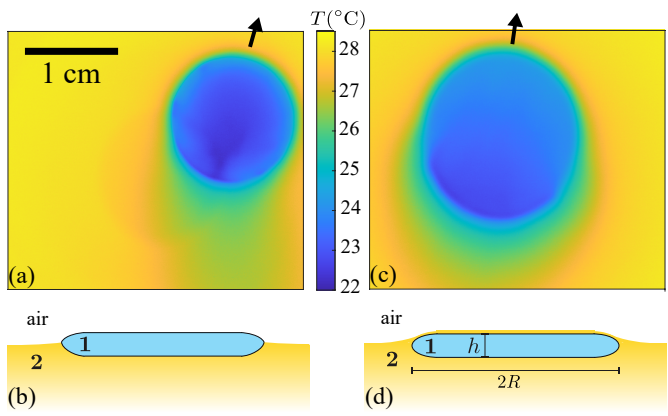}
\caption{Influence of the coating film on the temperature field of the drop surface. (a,b) Drop without film $S_{21}<0$. (c,d) Drop with a film $S_{21}>0$. In (a) the drop exhibits fluctuating cold patches. In (b) the drop exhibits a single cold patch confined to its aft. (a,c) Thermal images. (b,d) Schematics (not to scale). }
\label{fig_supmat}
\end{figure}

We observe that the temperature field on the surface of a drop without coating film fluctuates and exhibits cold patches appearing erratically (Fig. \ref{fig_supmat}(a)). These thermo-capillary instabilities are well known in evaporating sessile drops \cite{sefiane2008}. The erratic nature of the drop dynamics in this archetype results from these temperature fluctuations. When the coating film is present (Fig. \ref{fig_supmat}(c,d)) temperature fluctuations are absent. In this case, a cold patch appears on one side of the drop and finally results into steady thermocapillary stresses and dynamics (Fig. \ref{fig_supmat}(c)).  

\end{document}